\documentclass[english]{elsarticle}
\usepackage[T1]{fontenc}
\usepackage[latin9]{inputenc}
\usepackage{units}
\usepackage{textcomp}
\usepackage{amsmath}
\usepackage{amssymb}
\usepackage{graphicx}
\usepackage{esint}

\makeatletter

\newcommand{\lyxmathsym}[1]{\ifmmode\begingroup\def\b@ld{bold}
  \text{\ifx\math@version\b@ld\bfseries\fi#1}\endgroup\else#1\fi}

\providecommand{\tabularnewline}{\\}

\usepackage{a4wide}

\makeatother

\usepackage{babel}
\begin{document}

\title{Nonextensive lattice gauge theories: algorithms and methods}

\author{Rafael B. Frigori}

\address{Universidade Tecnológica Federal do Paraná (UTFPR). \\
Rua Cristo rei 19, CEP 85902-490, Toledo (PR), Brazil}

\ead{frigori@utfpr.edu.br}
\begin{abstract}
High-energy phenomena presenting strong dynamical correlations, long-range
interactions and microscopic memory effects are well described by
nonextensive versions of the canonical Boltzmann-Gibbs statistical
mechanics. After a brief theoretical review, we introduce a class
of generalized heat-bath algorithms that enable Monte Carlo lattice
simulations of gauge fields on the nonextensive statistical ensemble
of Tsallis. The algorithmic performance is evaluated as a function
of the Tsallis parameter $q$ in equilibrium and nonequilibrium setups.
Then, we revisit short-time dynamic techniques, which in contrast
to usual simulations in equilibrium present negligible finite-size
effects and no critical slowing down. As an application, we investigate
the short-time critical behaviour of the nonextensive hot Yang-Mills
theory at $q-$values obtained from heavy-ion collision experiments.
Our results imply that, when the equivalence of statistical ensembles
is obeyed, the long-standing universality arguments relating gauge
theories and spin systems hold also for the nonextensive framework. \end{abstract}
\maketitle

\section{Introduction}

There is increasing evidence that generalizations of the canonical
thermostatistics of Boltzmann-Gibbs (BG) are usefull to describe important
phenomenological aspects of relativistic hadronic collisions \cite{PHENIX_AuAu,q-scalling}.
Traditionally, a QCD inspired formula à la Hagedorn \cite{Hagedorn}
is employed to fit the cross sections $\left(\sigma\right)$ of hadrons
as a function of their transverse momenta $\left(p_{T}\right)$
\begin{equation}
E\frac{d^{3}\sigma}{d^{3}p}=C\cdot\left(1+\frac{p_{T}}{p_{0}}\right)^{-\alpha}\rightarrow\left\{ \begin{array}{c}
p_{T}\rightarrow0\Longrightarrow\exp\left(-\nicefrac{\alpha p_{T}}{p_{0}}\right)\\
p_{T}\rightarrow\infty\Longrightarrow\left(\frac{p_{T}}{p_{0}}\right)^{\alpha}
\end{array}\right.\label{Hagedorn}
\end{equation}
with parameters $C,$ $p_{0}$ and $\alpha.$ Where the mean transverse
momentum of the system $\left\langle p_{T}\right\rangle $ is related
to its hadronization temperature $T$ in equilibrium. 

However, such temperature shall naturally fluctuate among events \cite{Power_laws_in_Heavy_Ions},
in a clear far-from-equilibrium scenario. Therefore, the usual BG
picture shall be generalized to naturally accommodate such fluctuations,
this is done by considering a Tsallis \cite{Tsallis_original,Tsallis_Book}
distribution
\begin{equation}
E\frac{d^{3}\sigma}{d^{3}p}=C_{q}\cdot\left[1-\left(1-q\right)\frac{p_{T}}{T}\right]^{\frac{1}{1-q}}.\label{Hagedorn_Tsallis}
\end{equation}
Here $\alpha=\frac{1}{q-1},$ $p_{0}=\frac{T}{q-1}$ and $C_{q}$
is a normalization, whereas the nonextensive parameter $q$ is related
\cite{Interpretation_of_q_in_Appl} to the variance of $T$ by 
\begin{equation}
q=1+\frac{Var\left(T\right)}{\left\langle T\right\rangle ^{2}}.\label{q_Tsallis_var-T}
\end{equation}

Along the last years variations of the approach in Eq.(\ref{Hagedorn_Tsallis})
have been verified by different collaborations, as ALICE \cite{ALICE_},
ATLAS \cite{ATLAS} and CMS \cite{CMS} at LHC and PHENIX \cite{PHENIX}
and STAR \cite{STAR} at RHIC, which has fit experimental data by
power-like (Levy) distributions using Tsallis formulae \cite{Multiplicities_distribution_Tsallis,Species_examination_Tsallis}.

While equations Eq.(\ref{Hagedorn}) and Eq.(\ref{Hagedorn_Tsallis})
may seem similar from the mathematical point, their underlying physics
is quite distinct. The nonextensive expression admits a complex (steady
state) thermal equilibrium for any $p_{T},$ which can be described
by just two parameters $T$ and $q.$ Thus, it is an unifying statistical
mechanics approach that does not relly on any particular model, or
theoretical regime of a more fundamental theory (i.e. perturbative
vs nonperturbative QCD) \cite{q-scalling}, to be derived from.

Still, this conceptual difference plays a central role when modeling
heavy-ion collisions through a hydrodynamical approach with evolution
\cite{QGP_Hydro}. There, the transverse momentum distributions of
multiple particles species is usually described by a Boltzmann-Gibbs
Blast-Wave (BGBW) model, see \cite{TBW} (and references therein).
This brings most of physical insights about the behaviour of the fireball.
Nevertheless, such an equilibrium description is believed to break
at high $p_{T},$ when nonequilibrium effects and hard processes will
exhibit power-law tail \cite{Interpretation_of_q_in_Appl}. So, generalizations
of BGBW incorporating principles of Tsallis thermostatistics \cite{TBW}
are necessary (TBW), and in fact they have shown to be powerful enough
to describe experimental data \cite{PHENIX_AuAu,PHENIX,Experiments_TBW}.

While nonextensive extensions of well-known phenomenological models
is an atractive area, with potential implications \cite{Beck_nonextensive_SM_of_particles}
-- see also \cite{Walecka_Tsallis} for q-Walecka and \cite{NJL_Tsallis}
for q-NJL --- their derivation from first principles is not fully
understood yet \cite{Tsallis_PT_from_pQCD,Kohyama}. Therefore, it
is of high theoretical interest to generalize first-principle nonperturbative
methods, as the lattice formalism of nonabelian gauge theories \cite{LatticeQCD},
to the Tsallis ensemble. Moreover, from a pure computational perspective,
lattice simulations may benefit from Tsallis weight, hence it enhances
the tunneling rate among metastable states during phase transitions
\cite{Takaishi}. 

In this context, following the generalized master equation approach
of \cite{Drugo_qMetropolis_Ising}, we introduce a generalized hybrid
heat-bath algorithm to enable Monte Carlo simulations of lattice Yang-Mills
(YM) theory in the Tsallis ensemble. In addition, this algorithm can
be also easily adapted to other $SU\left(N\right)$ gauge theories.
Thus, we perform a rigorous analysis of algorithmic performance in
2d lattices, where the $SU\left(2\right)$ theory is exactly solvable.
This solution helps on evaluating integrated correlation-times of
critical plaquettes, as a function of the lattice-side and $q,$ while
investigating a region of \textit{constant physics} \cite{MHB}.
In agreement with previous studies \cite{Takaishi,Morishita-Johal_EGE-MUCA-Tsallis}
we observe that setups with $q>1$ induce significant improvements
on computational efficiency when compared to the canonical case $\left(q=1\right).$ 

On the other hand, when considering finite-temperature simulations
on equilibrium, universality has been a cornerstone principle to understand
the thermodynamics of gauge theories in the canonical ensemble. For
instance, even dynamical aspects of such theories, as their screening-mass
spectra \cite{Screening_masses}, were predicted from condensed-matter
analogous. Also, based on arguments of symmetry, QCD with two dynamic
quarks undergoes a phase transition with universal critical scaling
in the class of the 3d $O\left(4\right)$ continuous-spin model%
\footnote{Incidentally, this spin system can be simulated using a heat-bath
algorithm shared by lattice Y.M. theory \cite{MHB}%
} \cite{O4_scaling}. Furthermore, there is the long-standing argument
by Svetitsky and Yaffe \cite{Svet&Yaffe} relating critical quenched
$SU\left(N\right)$ gauge fields in $d+1$ dimensions to $Z_{N}$
spin systems in $d-$dim. 

Those simulations are challenging, not only because finite-size (FS)
effects --- and their necessary scaling extrapolations --- have to
be keep under control, but also because the well-know critical slowing
down (CSD) effect \cite{Sokal}. This implies exponentially diverging
correlation times of observables, and so their statistical errors.
A way to aleviate that computational burden comes from short-time
dynamical simulations, for a review see \cite{B.Zheng_review_short_time,Okano-Otobe-Zheng}.
This technique allows for extracting the critical behaviour, summarized
in a set of dynamic and static exponents, of spin-systems or gauge
fields without appreciable FS or CSD effects. This feature is rooted
on the findings \cite{Fisher-Halperin-etc} that even during a short-time
transient regime, before (Monte Carlo) equilibration happens, the
hamiltonian dynamics already exhibits universal scaling. 

Considering that few is known about the aforementioned critical properties
of nonextensive gauge theories, we investigate through short-time
simulations the finite-temperature $3d$ $SU\left(2\right)$ lattice
Y.M. theory $\left(YM_{2}^{3d}\right)$ in the Tsallis ensemble. Despite
of being considerably simpler than unquenched QCD, the $YM_{2}^{3d}$
theory is nontrivial. Actually, it has been shown to be a good theoretical
model for understanding fundamental properties of confinement. Concerning
gluonic propagators, no relevant discrepancies to QCD were found at
gauge groups \cite{SU2_meets_SU3} or dimensionality \cite{Whats_up_propagator}
levels. In addition, around criticality $YM_{2}^{3d}$ is related
to the bidimensional Ising model, an exactly solvable system, which
turns that gauge theory more auspicious for high-precision comparative
studies.

As a matter of fact, we have focused our simulations on values of
the nonextensive parameter $q\approx1$ --- as a perturbation around
BG thermodynamics --- and $q=1.10,$ a value favoured by experimental
data fits \cite{PHENIX_AuAu,q-scalling,Multiplicities_distribution_Tsallis}.
We have observed that for $q\approx1$ small deviations from usual
BG behaviour are seen. While in the $q>1$ regime the temperature
of the phase-transition is monotonically increased with $q$ (i.e.
$T_{q>1}^{crit.}>T_{q=1}^{crit.}$), as theoretically expected \cite{q-scalling,Interpretation_of_q_in_Appl}.
Besides that, by performing a Binder cumulant analysis (in equilibrium)
\cite{Landau-Binder} we confirm that our results are not afflicted
by any FS effect. More interestingly, not only the static and dynamic
exponents of the nonextensive theory, but also its universal cumulant
values, can be explained by (and generalizes) universality arguments
\cite{Svet&Yaffe}. 

The article is organized as follows: in Section 2 the nonextensive
thermostatistics of Tsallis is outlined. Its connections with the
usual Boltzmann-Gibbs statistics are discussed in the sense of superestatistics,
and finally, applications to gauge theories are provided. The Section
3 reviews short-time dynamic simulation techniques for gauge theories.
It gives an outlook on how to overcome the critical slowing down phenomena,
while evaluating static and dynamic exponents. Our generalized algorithmic
proposal is presented in Section 4, after briefly reviewing the theory
of Markov processes and (generalized) detailed balance. The necessary
modifications to usual heath-bath updating engines \cite{MHB} is
theoretically motivated and implemented. In Section 5, numerical results
on algorithmic performance are analysed for the 2d $SU\left(2\right)$
gauge theory and, the nonextensive relaxation dynamics for finite-temperature
3d $SU\left(2\right)$ theory is studied. Main conclusions and prospective
research directions are the focus of Section 6.

\section{Nonextensive thermostatistics of lattice gauge theories}

The lattice gauge theory formalism allows for \textit{ab initio} thermodynamic
analysis of quantum fields at finite-temperature nonperturbative regimes
\cite{LatticeQCD}. Most times it is performed in the quenched approximation,
where quark-loop effects are neglected. Within this approach the deconfinement
phase transition of $SU\left(N\right)$ theories can be related by
universality arguments to the magnetic transition of $Z_{N}$ spin
models \cite{Svet&Yaffe}.

A realization for pure gauge $SU\left(N\right)$ theories in $d-$dimensional
lattices is given \cite{LatticeQCD} by the Wilson action 
\begin{equation}
S_{W}\left[U\right]\equiv\beta\sum_{x}\sum_{\mu,\nu=1}^{d}\left\{ 1-\frac{1}{N}Re(TrP_{\mu\nu})\right\} ,\label{Wilson_Action}
\end{equation}
where gauge links $U_{\mu}\left(x\right)\in SU(N)$ are combined to
build a gauge-invariant plaquette
\begin{equation}
P_{\mu\nu}\equiv U_{\mu}\left(x\right)U_{\nu}\left(x+\hat{\mu}a\right)U_{\mu}^{-1}\left(x+\hat{\nu}a\right)U_{\nu}^{-1}\left(x\right).\label{Plaquete}
\end{equation}
The lattice-coupling $\beta=2N/g_{s}^{2}a^{4-d}$ is set in terms
of the gauge-field coupling $g_{s}$ and the physical lattice spacing
$a$. 

In the canonical ensemble the temperature of equilibrium is identified
with the inverse length of the temporal direction (i.e. $T^{-1}=a\cdot L_{t}$)
of an assymetric lattice, whose volume is $V=a^{d}\cdot L_{s}^{d-1}\cdot L_{t}$
\cite{LatticeQCD}. Thus, thermal expectation values of any gauge-invariant
operator $\mathcal{O}$ may be computed by
\begin{equation}
\left\langle \mathcal{O}\right\rangle _{BG}=\frac{\sum_{U}\mathcal{O}\left(U\right)e^{-S_{W}\left(U\right)}}{\sum_{U}e^{-S_{W}\left(U\right)}}.\label{Thermal_average}
\end{equation}
Among such observables is the (spatially averaged) Polyakov loop $\bar{W}\equiv\left\langle W\left(x,y,z\right)\right\rangle =\left\langle Tr\prod_{n=1}^{n=L_{t}}U_{t}\left(x,y,z,an\right)\right\rangle $,
the order parameter of deconfinement phase transition.

Tsallis introduced a nonextensive generalization of the usual canonical
ensemble \cite{Tsallis_original} by postulating a pseudo-additive
entropy 
\begin{equation}
S_{q}=\frac{\left(1-Tr\hat{\rho}^{q}\right)}{q-1}.\label{S_Tsallis}
\end{equation}
Here, the real-number $q$ (Tsallis parameter) regulates the degree
of non-additivity of the generalized entropy $S_{q}^{A+B}=S_{q}^{A}+S_{q}^{B}+(1-q)S_{q}^{A}S_{q}^{B},$
and $\hat{\rho}$ is a density operator. The explict form of $\hat{\rho}$
is obtained by a constrained maximization of $S_{q}$ \cite{Tsallis_Book,3-choices-of_Sq,Equivalent_4-versions_of_Tsallis_statistics}
which, for instance, leads to
\begin{equation}
\hat{\rho}=Z_{q}^{-1}e_{q}^{-\hat{H}/T}=Z_{q}^{-1}\left[1-\left(1-q\right)\hat{H}/T\right]^{1/\left(1-q\right)}.\label{Density_q-operator}
\end{equation}
Where $Z_{q}$ stands for the $q-$dependent partition function of
Tsallis (i.e. a normalization factor) and $\hat{H}$ is the Hamiltonian
of the system at physical temperature $T$. 

The resulting Tsallis statistics is also known to be a particular
case of superstatistics \cite{Beck_superestatistics}, derived as
a superposition of different BG statistics, with special relevance
for nonequilibrium systems. Thereby, the Tsallis weight $\omega_{k}$
can be obtained from a (Gamma) integral-transform over Boltzmann-Gibbs
weights 
\begin{equation}
\omega_{k}=\frac{1}{Z_{c}}\intop_{0}^{\infty}d\theta w_{c}\left(\theta\right)e^{-\theta\beta E_{k}},\label{Tsallis-weight_from_BG}
\end{equation}
where $q=1+1/c$ and $w_{c}\left(\theta\right)=\frac{c^{c}}{\Gamma\left(c\right)}\theta^{c-1}e^{-c\theta}.$
Thus, any BG expectation value can be converted into a Tsallis one
if it is known as a function of $\beta.$ In particular, the respective
partition functions of Boltzmann-Gibbs $\left(Z_{BG}\right)$ and
of Tsallis $\left(Z_{T}\right)$ are related by
\begin{equation}
Z_{T}\left(\beta\right)=\sum_{k}\intop_{0}^{\infty}d\theta w_{c}\left(\theta\right)e^{-\theta\beta E_{k}}=\intop_{0}^{\infty}d\theta w_{c}\left(\theta\right)Z_{BG}\left(\theta\beta\right).\label{Z_BG-Z_TS}
\end{equation}

Considering the case of pure gauge theories, where $Z_{BG}=\int DUe^{-S_{W}\left(U\right)}$,
the expression in Eq.(\ref{Z_BG-Z_TS}) is explicitly written as
\begin{equation}
Z_{T}=\frac{c^{c}}{\Gamma\left(c\right)}\intop_{0}^{\infty}d\theta e^{-c\theta}\theta^{c-1}\int DUe^{-S_{W,\tilde{\beta}}\left(U\right)},\label{ZTS_from_ZBG}
\end{equation}
where $S_{W,\tilde{\beta}}\left(U\right)$ is the usual Wilson action
of Eq.(\ref{Wilson_Action}) evaluated%
\footnote{In \cite{T_V_fluct_equivalence} (and references therein) it was
shown that, for systems with constant total energy, volume fluctuations
are equivalent to temperature fluctuations. In fact, both these (Gamma)
fluctuations can equivalently lead to the Tsallis form of the respective
distributions for energy spectra. Thus, it is quite natural to employ
a symmetric lattice-coupling $\beta$ (i.e. the same $\beta$) for
space and ``time'' directions in the Wilson action.%
} for $\tilde{\beta}=\theta\beta.$ Then, by assuming a finite $c$,
the $\theta-$integration and the path-integral can be exchanged 
\begin{equation}
Z_{T}=\int DU\frac{c^{c}}{\Gamma\left(c\right)}\intop_{0}^{\infty}d\theta e^{-c\theta}\theta^{c-1}e^{-S_{W,\tilde{\beta}}\left(U\right)}\rightarrow Z_{q}=\int DUe_{q}^{-S_{W}\left(U\right)}.\label{ZTS_from_ZBG_LGTs}
\end{equation}

Therefore, once the Tsallis formalism reduces to the usual BG approach
in the limit $q=1$ (i.e. $S_{q=1}=S_{BG}=-Tr\hat{\rho}\ln\hat{\rho}$)
\cite{Tsallis_Book,Kohyama}, a $q-$expectation value $\left\langle \cdot\right\rangle _{q}$
that generalizes Eq.(\ref{Thermal_average}) may be written \cite{Kohyama,Equivalent_4-versions_of_Tsallis_statistics}
as
\begin{equation}
\left\langle \mathcal{O}\right\rangle _{q}=\frac{\sum_{U}\mathcal{O}\left(U\right)e_{q}^{-S_{W}\left(U\right)}}{\sum_{U}e_{q}^{-S_{W}\left(U\right)}}.\label{q-average}
\end{equation}

\section{Short-time critical dynamics}

Renormalization group techniques predict \cite{Fisher-Halperin-etc}
that after a sudden quench to the critical temperature $T_{c}$ many
physical systems can display universal dynamical behaviour even during
the early (nonequilibrium) evolution times. Curiously, along this
transient process finite-size effects and critical slowing down phenomena
\cite{Sokal} are almost absent. This may be understood by realizing
that in such simulations observables are averaged over time-slices
from independent Markov chains, which are started from similar initial
states \cite{Landau-Binder}. 

Thence, this technique allows for efficient characterization of critical
properties of systems undergoing relaxation to thermal equilibrium.
For instance, correlation scales and critical exponents may be extracted
by studying the dynamic evolution of appropriate functions of the
order parameters. In particular, the order parameter for gauge theories
is the so-called Polyakov loop $\left(W_{\vec{r}}\right),$ which
seems an \textit{effective magnetization} $M$ of spin systems \cite{Okano-Otobe-Zheng}.
Its (time-dependent) definition is given by
\begin{equation}
M\left(t\right)\doteq\left\langle \frac{1}{L_{s}^{d-1}}\sum_{\vec{r}}W_{\vec{r}}\left[t\right]\right\rangle _{samples},\label{M_Poly}
\end{equation}
where $\left\langle \cdots\right\rangle _{samples}$ denotes averaging
over configurations at the same Monte Carlo instant $\left[t\right]$. 

When considering the dynamic relaxation from a completely ordered
state, i.e. with initial magnetization $m_{0}=1$, a general scaling
form for the $k-th$ moment of the magnetization $M$ emerges 
\begin{equation}
M^{(k)}(t,\tau,L_{s})=b^{\lyxmathsym{\textminus}k\beta/\nu}M^{(k)}(b^{\lyxmathsym{\textminus}z}t,b^{1/\nu}\tau,b^{\lyxmathsym{\textminus}1}L_{s}).\label{kth-moment-of-M}
\end{equation}
Here $t$ is the MC time of the dynamic relaxation, $\tau$ is the
reduced coupling constant, $b$ is a rescaling factor, $\beta/\nu$
is the ratio between two (static) critical exponents, $z$ is a dynamic
exponent and $L_{s}$ is the lattice side. This scaling form has been
shown to be valid in the short-time regime for a number of different
physical systems including gauge theories \cite{Okano-Otobe-Zheng,Jaster_HMC}. 

By choosing $b=t^{1/z}$ as the rescaling factor in Eq.(\ref{kth-moment-of-M})
and assuring that $\tau=0$, it leads to a power-law behaviour for
the magnetization $\left(i.e.\; k=1\right)$ given by
\begin{equation}
M\left(t\right)\sim t^{-\beta/\nu z}.\label{M_t_scaling}
\end{equation}
In addition, the scaling of the cumulant 
\begin{equation}
U=\frac{M^{2}\left(t\right)}{M\left(t\right)^{2}}-1\label{U_t}
\end{equation}
can be expressed \cite{Jaster_HMC} in terms of the space dimension
$d$ as 
\begin{equation}
U\left(t\right)\sim t^{d/z},\label{U_T_scaling}
\end{equation}
thus providing the value of $z$ while fixing the ratio $\beta/\nu.$ 

Furthermore, at the critical line the autocorrelation of the order
parameter

\begin{equation}
A\left(t\right)\doteq\left\langle \frac{1}{L_{s}^{2\left(d-1\right)}}\sum_{\vec{r}}W_{\vec{r}}\left[t\right]W_{\vec{r}}\left[0\right]\right\rangle _{samples},\label{A_Poly}
\end{equation}
also obeys a power law $A\left(t\right)\sim t^{-\eta/2z},$ while
in the low temperature phase%
\footnote{It is worth to mention that when considering spin systems this relation
is valid for the high-temperature phase, as explained by universal
mappings described in \cite{Svet&Yaffe}. %
} it is described by the ansatz 
\begin{equation}
A\left(t\right)\sim t^{-\eta/2z}\exp\left(-t/\xi_{t}\right).\label{A_Poly_scaling}
\end{equation}
Where the nonequilibrium autocorrelation time $\xi_{t}$ is related
to the equilibrium autocorrelation length $\xi_{s}$ through $\xi_{t}\propto\xi_{s}^{z}$
\cite{Jaster_XY_DTC}.

\section{A generalized heat-bath algorithm for the Tsallis ensemble}

Dynamical Monte Carlo simulations use Markov chains designed to generate,
when in equilibrium, a desired target probability distribution $P\left(E\right).$
To ensure this, a sufficient condition is known to be the detailed
balance 
\begin{equation}
\omega\left[g\rightarrow g'\right]P\left[E\left(g\right)\right]-\omega\left[g'\rightarrow g\right]P\left[E\left(g'\right)\right]=0.\label{Usual_detailed_balance}
\end{equation}
Where, $\omega\left[g'\rightarrow g\right]$ is the transition-rate
of the system configuration from $g$ to $g',$ and $E\left(g\right)$
{[}$E\left(g'\right)${]} is the energy --- or alternatively, the
action $S\left(g\right)$ --- of the system before {[}after{]} the
transition \cite{LatticeQCD}. 

Different updating algorithms implement Eq.(\ref{Usual_detailed_balance})
by constructing particular transition rules. For instance, a new configuration
$g'$ can be proposed to replace $g$ with an \textit{a priori} selection
probability $p_{T,g}\left(g'\right)$ \cite{Creutz_OVR}. After that,
the proposal may be accepted with a given conditional probability
$P_{A}$ satisfying Eq.(\ref{Usual_detailed_balance}). A realization
of this last step is given by the general Metropolis choice 
\begin{equation}
P_{A}=\min\left\{ 1,\frac{p_{T,g'}\left(g\right)\times P\left[E\left(g'\right)\right]}{p_{T,g}\left(g'\right)\times P\left[E\left(g\right)\right]}\right\} .\label{Transition-rate_Metropolis-Hastings}
\end{equation}

In particular, when $p_{T,g}\left(g'\right)=p_{T,g'}\left(g\right)$
--- and $P\left(E\right)$ satisfies the BG statistics --- the acceptance
on Eq.(\ref{Transition-rate_Metropolis-Hastings}) reduces to the
well-known Metropolis criterion
\begin{equation}
\omega\left[g\rightarrow g'\right]=\min\left\{ 1,\exp\left[-\beta\left(E\left(g'\right)-E\left(g\right)\right)\right]\right\} .\label{Transition_usual-Metropolis}
\end{equation}
Alternatively, for local actions, one can choose $g'$ with probability
$p_{T,g}\left(g'\right)\propto\exp\left[-\beta E\left(g'\right)\right]$
to obtain the heat-bath algorithm \cite{Creutz_OVR}, whose $P_{A}=1$.

Fortunately, for pure $SU\left(2\right)$ gauge theories, the Wilson
action Eq.(\ref{Wilson_Action}) enables an exact implementation of
the heat-bath algorithm --- i.e., by taking $p_{T,U_{\mu}}\left(U_{\mu}^{new}\right)\propto\exp\left[-S_{1-link}\left(U_{\mu}^{new}\right)\right]$
--- since $S_{W}$ can be expressed as a sum of single-link (local)
actions 

\begin{equation}
S_{1-link}=-\frac{\beta}{2}Tr\left[U_{\mu}\left(x\right)H_{\mu}\left(x\right)\right].\label{S_1link}
\end{equation}
Here the gauge link $U_{\mu}\left(x\right)\in SU\left(2\right),$
$H_{\mu}\left(x\right)$ is the sum of neighbour staples written as
$H_{\mu}\left(x\right)=N_{\mu}\left(x\right)\tilde{H_{\mu}}\left(x\right),$
with $\tilde{H_{\mu}}\left(x\right)\in SU\left(2\right)$ and $N_{\mu}\left(x\right)=\sqrt{\det H_{\mu}\left(x\right)}$.

Then, by imposing over Eq.(\ref{S_1link}) the invariance of group
measure one obtains \cite{MHB,Creutz} the update step 
\begin{equation}
U_{\mu}\left(x\right)\longrightarrow U_{\mu}^{new}\left(x\right)=V\tilde{H}_{\mu}^{\dagger}.\label{HB_update}
\end{equation}
Where the unimodular evolution matrix $V=v_{0}I+i\vec{\cdot v}\cdot\vec{\sigma}\in SU\left(2\right)$
is generated by randomly taking $v_{0}$ according to the distribution
\begin{equation}
P\left(v_{0}\right)\varpropto\sqrt{1-v_{0}{{}^2}}\exp\left(\beta Nv_{0}\right)dv_{0},\label{Pv0_Creutz}
\end{equation}
while the components of $\vec{v}$ are isotropically chosen from $\mathbb{R}^{3}.$ 

An implementation of Eq.(\ref{Pv0_Creutz}) was originally proposed
by Creutz \cite{Creutz}, it consists on directly sampling $v_{0}$
with probability $P\left(v_{0}\right)\sim\exp\left(\beta Nv_{0}\right)$
and so correcting for the $\sqrt{1-v_{0}\text{\texttwosuperior}}$
factor by rejection. Besides that, once invariance under group measure
does not constrain the vectorial part of the evolution matrix $V,$
a microcanonical overrelaxation step \cite{MHB} may be incorporated
by taking\textit{ $\vec{v}\rightarrow-sgn(\vec{v}\cdot\vec{w})\vec{v},$}
with $\mathrm{\mathcal{W}}=w_{0}I+i\cdot\vec{w}\cdot\vec{\sigma}=U_{\mu}^{old}\left(x\right)\tilde{H_{\mu}}\left(x\right)$
where $sgn$ denotes the sign function. Still, this algorithm (MHB
\cite{MHB}) can be iteratively applied for the $SU\left(2\right)$
subgroups of $SU\left(N\right),$ so producing a pseudo heat-bath
approach for any quenched lattice gauge theory \cite{LatticeQCD}.

However, in a more general context, as to simulate gauge theories
in the nonextensive ensemble of Tsallis, the target probabilities
$P\left(E\right)$ in Eq.(\ref{Usual_detailed_balance}) will become
a q-generalized statistical distributions $P_{q}\left(E\right)$ \cite{Equivalent_4-versions_of_Tsallis_statistics}.
As discussed in \cite{Drugo_qMetropolis_Ising}, in this case the
usual Metropolis updating scheme in Eq.(\ref{Transition_usual-Metropolis})
becomes nonlocal even for local actions. It comes from the fact that
q-exponential functions are non-additive, so single-link modifications
introduce changes in the system energy that is spread all over the
lattice. 

To circunvent such additional computational burden one would need
to devise a way to retrieve locality in the Monte Carlo updates. This
can be accomplished by algorithms derived from a q-generalized detailed
balance condition (see for details \cite{Drugo_qMetropolis_Ising},
and references therein) written%
\footnote{For the particular definition of $P_{q}\left(E\right)$ in \cite{Drugo_qMetropolis_Ising}
employing escort probabilities as prescribed by TMP convention \cite{Equivalent_4-versions_of_Tsallis_statistics}
one has to use slightly different definitions for $q-$operators$.$
On the other hand, we employ TO convention for $P_{q}\left(E\right)$
without need to escort probabilities.%
} as
\begin{equation}
\omega\left[g\rightarrow g'\right]\otimes_{q}P_{q}\left[E\left(g\right)\right]\ominus_{q}\omega\left[g'\rightarrow g\right]\otimes_{q}P_{q}\left[E\left(g'\right)\right]=0.\label{daSilva_gen_detailed_balance}
\end{equation}
Where use is made of the so-called \cite{Tsallis_Book} algebraic
q-operators
\begin{equation}
a\oplus_{q}b=a+b+\left(1-q\right)ab,\label{q_add}
\end{equation}
\begin{equation}
a\ominus_{q}b=\frac{a-b}{1+\left(1-q\right)b},\label{q_sub}
\end{equation}
\begin{equation}
a\otimes_{q}b=\left(a^{1-q}+b^{1-q}-1\right)^{1/\left(1-q\right)},\label{q_prod}
\end{equation}
\begin{equation}
a\oslash_{q}b=\left(a^{1-q}-b^{1-q}+1\right)^{1/\left(1-q\right)}.\label{q_div}
\end{equation}
Which recovers the additive property of the argument $\exp_{q}\left(a\right)\exp_{q}\left(b\right)=\exp_{q}\left(a\oplus_{q}b\right)$
as well as $\exp_{q}\left(a\right)/\exp_{q}\left(b\right)=\exp_{q}\left(a\ominus_{q}b\right)$
while $\exp_{q}\left(a\right)\otimes_{q}\exp_{q}\left(b\right)=\exp_{q}\left(a+b\right)$
and $\exp_{q}\left(a\right)\oslash_{q}\exp_{q}\left(b\right)=\exp_{q}\left(a-b\right).$ 

An imediate solution of Eq.(\ref{daSilva_gen_detailed_balance}),
for systems with local actions, was given in \cite{Drugo_qMetropolis_Ising}
as a generalized Metropolis algorithm (q-Metropolis)
\begin{equation}
\omega\left[g\rightarrow g'\right]=\min\left\{ 1,\exp_{q}\left[-\beta\left(E\left(g'\right)-E\left(g\right)\right)\right]\right\} ,\label{daSilva_gen_Metropolis}
\end{equation}
whose transitions depend only on the energy difference between the
updated site and its neighbours. 

Besides that, when considering pure gauge theories, another natural
solution for Eq.(\ref{daSilva_gen_detailed_balance}) is a q-generalized
heat-bath algorithm (q-MHB) with \textit{a priori} probabilities given
by $p_{T,U_{\mu}}\left(U_{\mu}^{new}\right)\propto\exp_{q}\left[-S_{1-link}\left(U_{\mu}^{new}\right)\right].$
This algorithm satisfies a generalization (with q-operators) of Eq.(\ref{Transition-rate_Metropolis-Hastings})
--- derivable from Eq.(\ref{daSilva_gen_detailed_balance}) --- as
does q-Metropolis, whose large-repetition limit matches q-MHB \cite{Drugo_qMetropolis_Ising,Creutz_OVR}.
A straightforward implementation comes from modifying only the single-link
update step of usual MHB%
\footnote{Hence the probability density $P_{q}\left(v_{0}\right)$ can be generated
by rejection from $p_{q}\left(v_{0}\right)\sim exp_{q}\left(\beta Nv_{0}\right),$
one shall obtain $p_{q}\left(v_{0}\right)$ by the transform method
\cite{Numerical_Recipes}. For instance, $x$ is randomly drawn following
a general distribution as $p_{q}\left(x\right)\sim exp_{q}(c\cdot x)$
by computing $x=-q'ln_{q'}\left(U\right)/c$, whereas the random $U\in\left(0,1\right),$
while $q'=\left(2-q\right)^{-1}$ and $ln_{q'}\left(U\right)=\left(U^{1-q'}-1\right)/\left(1-q'\right).$%
} in Eq.(\ref{Pv0_Creutz}) to
\begin{equation}
P\left(v_{0}\right)\rightarrow P_{q}\left(v_{0}\right)\varpropto\sqrt{1-v_{0}\text{\texttwosuperior}}\exp_{q}\left(\beta Nv_{0}\right)dv_{0}.\label{P_1-link_qMHB}
\end{equation}

\section{Numerical results}

\subsection*{Algorithmic performance}

Whenever the equivalence of statistical ensembles holds \cite{Morishita-Johal_EGE-MUCA-Tsallis}
reweighting methods \cite{Landau-Binder} allows for converting thermal
averages among different ensembles. For instance Eq.(\ref{Thermal_average})
and Eq.(\ref{q-average}) may be related \cite{Takaishi} by
\begin{equation}
\left\langle \mathcal{O}\right\rangle _{BG}=\left\langle \frac{\mathcal{O}\left(U\right)e^{-S_{W}\left(U\right)}}{e_{q}^{-S_{W}\left(U\right)}}\right\rangle _{TS}/\left\langle \frac{e^{-S_{W}\left(U\right)}}{e_{q}^{-S_{W}\left(U\right)}}\right\rangle _{TS}.\label{Reweighting}
\end{equation}
So, employing the Tsallis weight would be preferable than (and interchangeable
to) the Boltzmann one when simulations become more efficient in the
former ensemble.

\begin{figure}
\centering{}\includegraphics[width=7.5cm]{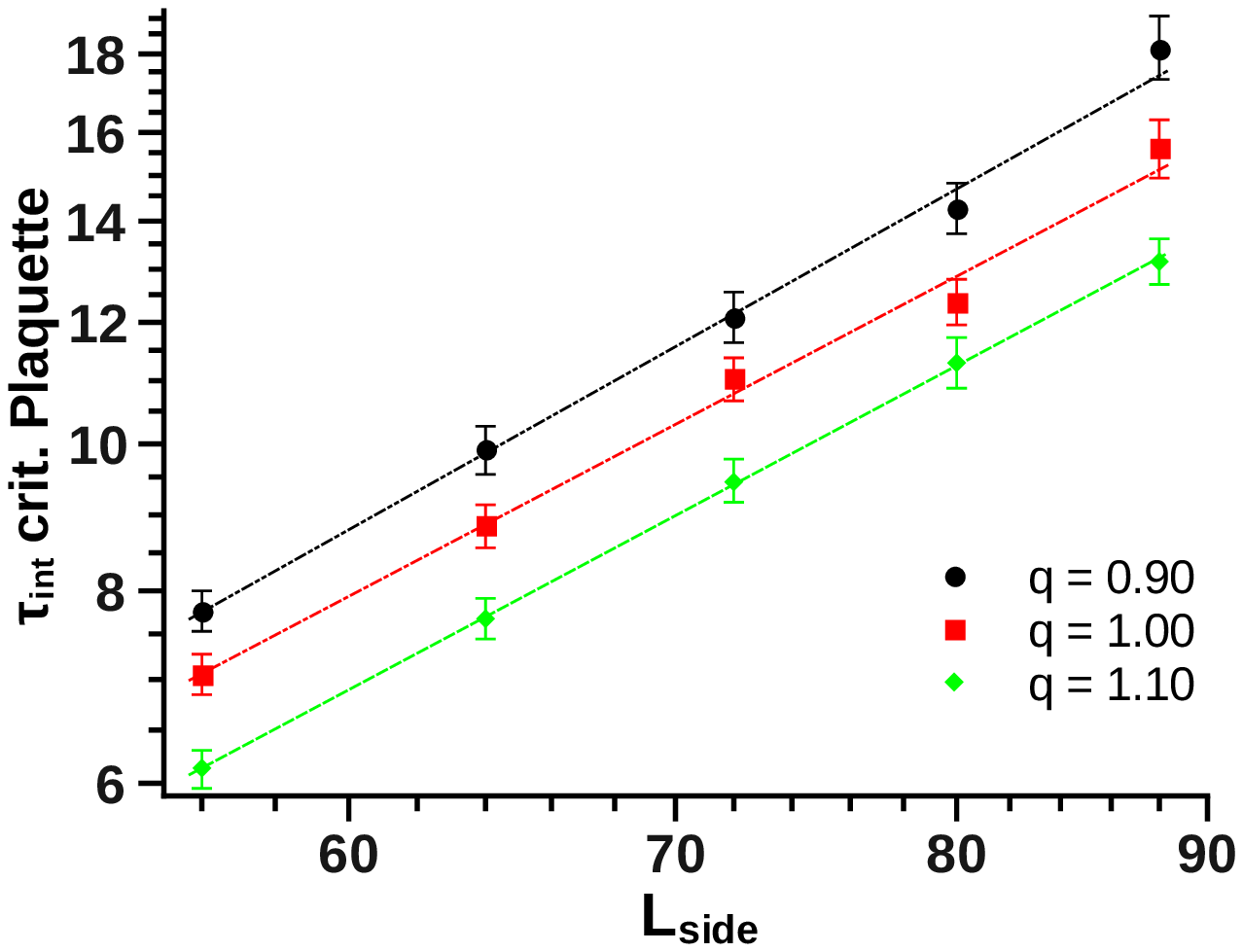}\includegraphics[width=7.5cm]{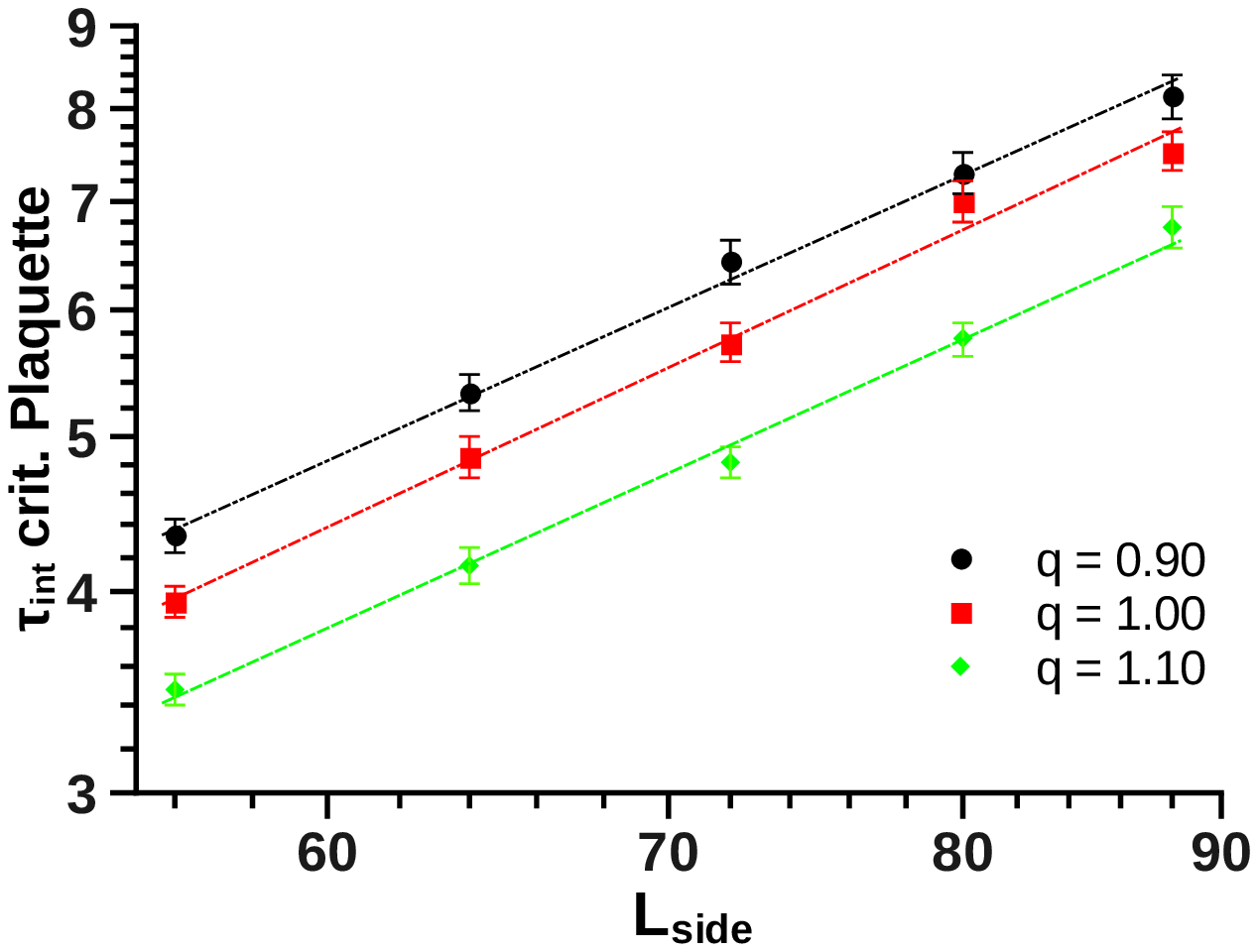}\caption{Integrated auto-correlation times of $M\times M$ plaquettes and their
fits to $\tau_{int}=a\cdot L_{side}^{z}$ for different lattice sides
and $q-$parameter values. {[}Left panel{]} results for the nonextensive
heat-bath algorithm (q-MHB) without overrelaxation. {[}Right panel{]}
the hybrid (overrelaxed) version of the same (q-MHB) algorithm.}
\end{figure}

Since statistical Monte Carlo errors are proportional to $\sqrt{2\tau_{int}},$
issues related to algorithmic efficiency may be set by computing the
integrated auto-correlation time 

\begin{equation}
\tau_{int}\left(\mathcal{O}\right)=\frac{1}{2}+\underset{t}{\sum}\rho_{f\mathcal{O}}\left(t\right).\label{Tau_Int}
\end{equation}
Where for a given physical observable $\mathcal{O}$ one defines $\rho_{f\mathcal{O}}=\frac{\left\langle \mathcal{O}_{i}\mathcal{O}_{i+t}\right\rangle -\left\langle \mathcal{O}_{i}\right\rangle ^{2}}{\left\langle \mathcal{O}_{i}^{2}\right\rangle -\left\langle \mathcal{O}_{i}\right\rangle ^{2}}$
\cite{Sokal}. Thereby, numerical errors in Eq.(\ref{Tau_Int}) can
be estimated by the Madras-Sokal formula \cite{Sokal} employing
self-consistent windowing \cite{MHB}. 

Generally a usual finite-size scalling $\tau_{int}\varpropto L_{side}^{\mathtt{\mathrm{\mathbf{z}}}}$
is expected%
\footnote{This $\mathcal{\mathsf{\mathtt{\mathrm{\mathbf{z}}}-}}$exponent is
not to be confused with the physical (dynamic) critical exponent $z$
measured by short-time relaxation techniques.%
}, and so the most efficient thermalization algorithm produces the
smallest $\mathtt{\mathrm{\mathbf{z}}}-$values for a set of observables.
Hence correlations increase with the lattice side, one supposes that
the best suited observables $\mathcal{O}$ for performance evaluations
are extended gauge-invariant quantities measured on regions of ``constant
physics''. This constraint may be ensured for instance by keeping
the ratio $\beta=L_{side}^{2}/32$ fixed. In particular, we considered
``critical'' plaquettes of $M\times M$ size, once $M=\sqrt{\frac{2\beta}{3}}\left(1+\frac{1}{4\beta}\right)$
scales with the correlation length $\xi$ of the 2d $SU(2)$ gauge
theory \cite{MHB}. 

The effects of tuning the nonextensive Tsallis parameter in the range
$0.9\leq q\leq1.10$ were investigated while lattice volumes were
set to $V=\left\{ 56^{2},64^{2},72^{2},80^{2},88^{2}\right\} .$ Our
q-generalized heat-bath algorithm and its overrelaxed version were
also compared for same volumes and $q$-values. The results obtained
after regression using $\tau_{int}=a\cdot L_{side}^{z}$, see Figure
(1), indicate that simulations with $q\gtrsim1$ are benefited by
the Tsallis approach which induces considerable decrease in $\tau_{int}.$
As a consequence, at largest volumes our simulations using $q=1.1$
are up to 9\% faster than the ones running under the usual (i.e. canonical)
setup%
\footnote{It deserves to be noted that tuning $q$ seems to just improve the
$a$ factor in $\tau_{int}=a\cdot L_{side}^{z},$ while the overrelaxation
has a stronger impact on $\mathcal{\mathsf{\mathtt{\mathrm{\mathbf{z}}.}}}$
Thus, the typical values found for heat-bath updates imply $\mathcal{\mathsf{\mathtt{\mathrm{\mathbf{z}}}}}\sim1.8(1)$
without using overrelaxation, and $\mathcal{\mathsf{\mathtt{\mathrm{\mathbf{z}}}}}\sim1.4(1)$
(for any $q$) when this microcanonical step is added. %
} at $q=1.0$.

\subsection*{Short-time dynamic simulations}

In this section we employ the previously described short-time dynamic
techniques to study the finite-temperature critical behaviour of $SU(2)$
lattice gauge theory in $3d.$ Our simulations were started from ordered
initial configurations with \textit{$m_{0}=1,$} which has been proven
to be an advantageous choice \cite{Jaster_XY_DTC}. 

\begin{figure}
\centering{}\includegraphics[width=7.5cm]{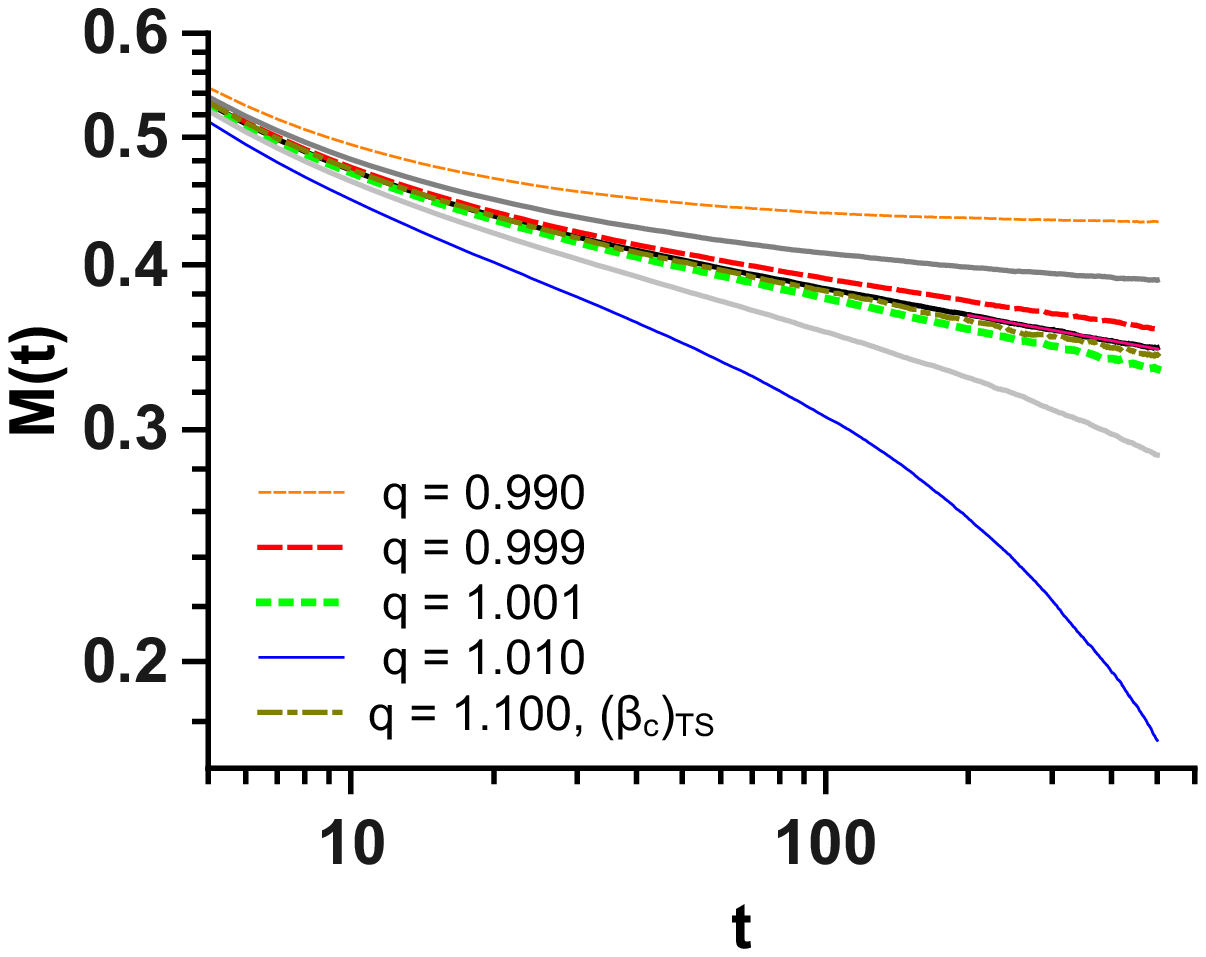}\includegraphics[width=7.5cm]{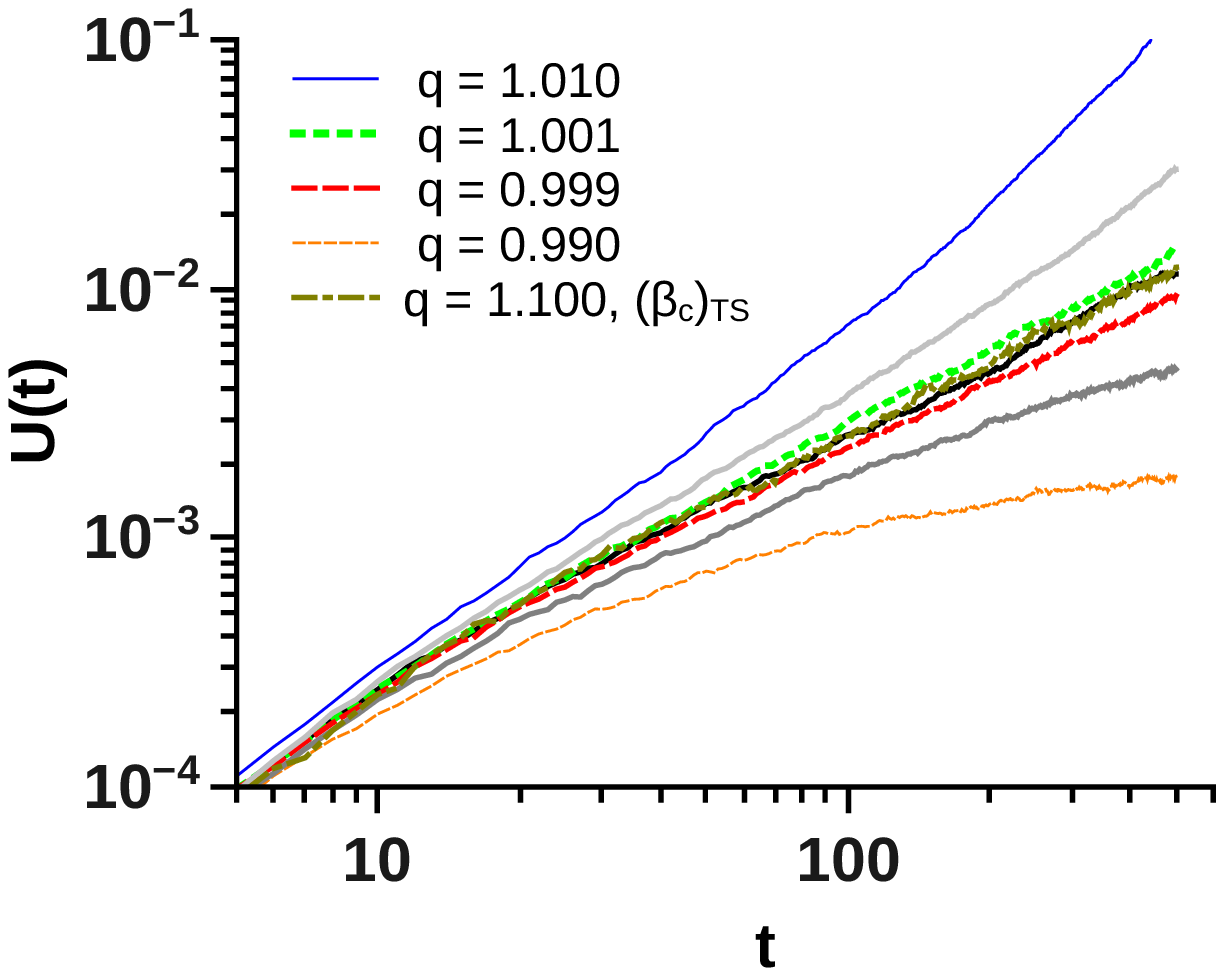}\caption{The dynamical evolution in Monte Carlo time of the Magnetization $M\left(t\right)$
{[}left panel{]} and of the cumulant $U\left(t\right)$ {[}right panel{]}.
Simulations were performed for different values of $q$ in the range
$\left[0.99,1.01\right]$ at the canonical critical lattice coupling
$\beta=\beta_{c}=3.4505.$ For comparative purposes canonical (i.e.
$q=1$) simulations at $\beta=\beta_{c}$ (bold black line), $\beta=0.99\cdot\beta_{c}$
(bold light gray) and $\beta=1.01\cdot\beta_{c}$ (bold dark gray)
are also ploted. In addition, there is a (bold dot dashed) curve at
$q=1.100$ which was tuned to the (shifted) critical coupling $\beta=\left(\beta_{c}\right)_{TS}$
in the Tsallis ensemble.}
\end{figure}

\begin{table}
\centering{}%
\begin{tabular}{|c|c|c|c|c|}
\hline 
$q$ & $z$ & $\beta/\nu$ & Start & Algorithm\tabularnewline
\hline 
\hline 
0.999 & 2.267(5) & 0.125(2) & cold & q-MHB\tabularnewline
\hline 
1.001 & 2.038(6) & 0.155(1) & cold & q-MHB\tabularnewline
\hline 
1.000 & 2.008(8) & 0.127(5) & cold & q-MHB\tabularnewline
\hline 
$1.10{}^{\dagger}$ & 2.139(9) & 0.124(4) & cold & q-MHB\tabularnewline
\hline 
$1.00^{*}$ & 2.155(3)  & 0.125 & hot & HB\tabularnewline
\hline 
\end{tabular}\caption{Static and dynamic critical exponents for the $3d$ $SU\left(2\right)$
pure gauge theory at its canonical critical lattice coupling $\beta\equiv\beta_{c}=3.4505,$
for different values of $q$ using our q-MHB algorithm, and cold starts.
The data in $\left(q=1.10{}^{\dagger}\right)$ was obtained by interpolation
at the nonextensively shifted critical coupling $\left(\beta=1.3275\cdot\beta_{c}\right)$.
For comparisons to predictions from universality, in $\left(q=1.00{}^{*}\right)$
it is shown results for the $2d$ critical Ising model simulated from
hot starts, and using heat-bath (HB) \cite{Okano-Otobe-Zheng}, in
the BG ensemble. }
\end{table}

For each value of the Tsallis parameter, taken in the vicinity of
the Boltzmannian limit $q\rightarrow1^{\pm}$, we have run 5000 simulations
initialized from different random seeds. The largest lattice volumes
we considered $V=128^{2}\times2$ allow for negligible finite-size
effects, which was also verified by a Binder cumulant analysis in
equilibrium, see below. Thus, relaxation was studied in those simulations
by evolving the system during 500 steps in Monte Carlo time where
the effects of nonextensivity on observables Eq.(\ref{M_Poly}) and
Eq.(\ref{U_t}) was monitored. The employed statistical error analysis
was standard, so data was grouped in independent blocks to compute
uncorrelated standard deviations \cite{DTC-Frigori}. 

The results for $M\left(t\right)$ and $U\left(t\right)$ are summarized
in Figure (2), which exhibits the outputs from simulations performed
at $q=\left\{ 0.990,0.999,1.000,1.001,1.010\right\} $ at canonical
critical coupling $\beta=\beta_{c}=3.4505.$ For comparative purposes
also the data from usual canonical simulations (i.e. at Boltzmaniann
limit $q=1$) with $\beta=1.01\cdot\beta_{c}$ (in bold dark gray),
$\beta=\beta_{c}$ (in bold black) and $\beta=0.99\cdot\beta_{c}$
(in bold light gray) are shown. Moreover, there are curves of $M\left(t\right)$
and $U\left(t\right)$ (dot-dashed dark yellow) evaluated at (best
approximation for) the shifted critical coupling $\beta_{TS}=1.325\cdot\beta_{c}$
for $q=1.10$ in the Tsallis ensemble. 

\begin{figure}
\centering{}\includegraphics[width=7.5cm]{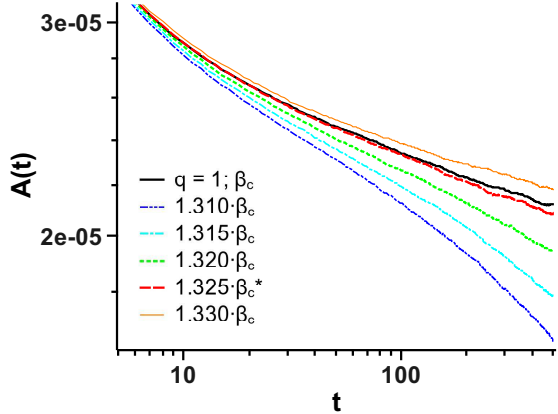}\caption{The dynamical evolution in Monte Carlo time of the auto-correlation
of the magnetization $A\left(t\right).$ The continuos bold black
line is the canonical BG simulation at critical point, i.e. $q=1$
and $\beta=\beta_{c}.$ The other curves were simulated at fixed $q=1.1$
and different values of lattice coupling $\beta>\beta_{c}.$}
\end{figure}

\begin{figure}
\centering{}\includegraphics[width=8cm]{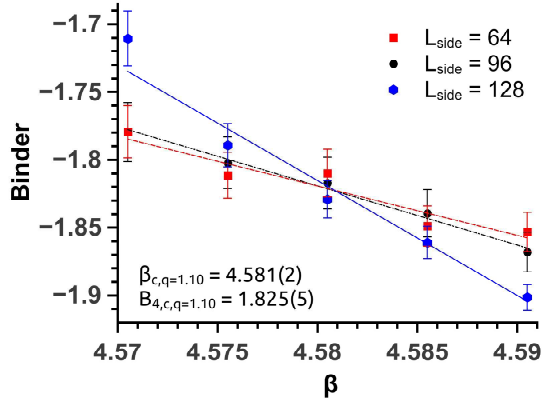}\caption{The fourth-order Binder cumulant of the Polyakov loop for $q=1.1$
and lattice sides $L=\left\{ 64,96,128\right\} $ as a function of
the lattice coupling $\beta.$ The crossing happens at the critical
coupling $\beta_{c}=\tilde{\beta}_{c,q=1.1}\approx4.581(2)$ (in the
Tsallis ensemble) where $B_{4}\approx1.825(5),$ so it is in nice
agreement with predictions from universality (see text).}
\end{figure}

Under close examination, it becomes clear that our data agrees with
the theoretical predictions \cite{q-scalling,Why_Tsallis_in_hQCD}
that increasing $q$ above the unit is analogous to decrease the temperature
of the system, while the converse effect is emulated by setting $q<1$.
Thereby, the critical exponents extracted from fits of Eq.(\ref{M_t_scaling})
and Eq.(\ref{U_T_scaling}) are compiled in Table (1). There one finds
that in a narrow range $0.999\leq q\leq1.001$ not only ensemble equivalence
\cite{Morishita-Johal_EGE-MUCA-Tsallis} but also universality arguments
\cite{Svet&Yaffe,Fisher-Halperin-etc} (approximately) hold when
comparing data from the $SU(2)$ theory and $Z_{2}$ spin-system.
Actually, one may suppose that an exact numerical match for such critical
exponents would just happen when the shifts on the critical lattice
coupling induced by nonextensive effects are properly considered.

To test that last hypothesis we have employed two different scaling
methods to get ``$\beta_{c}-$shifted'' (i.e., $\tilde{\beta}_{c,q}$)
with $q$ fixed around the phenomenologically motivated value $q=1.10$
\cite{q-scalling,PHENIX}. First, we performed a graphical matching
procedure by monitoring the autocorrelation of the order parameter
Eq.(\ref{A_Poly}) as a function of increasing lattice couplings $\beta,$
see Figure (3). Considering that an exact scalling law Eq.(\ref{A_Poly_scaling})
is well-known for such observable and, $\xi_{t}\rightarrow\infty$
at the critical point (deconfinement transition), we could locate
the nonextensively shifted critical coupling (by linear interpolation)
at $\tilde{\beta}_{c,q=1.1}\approx1.3275\cdot\beta_{c}\approx4.5805(8).$ 

The other approach is computationally more demanding, and complementar,
once it allows for locating eventual finite-size effects quite easily
\cite{Landau-Binder}. It consists on evaluating, by usual simulations
in equilibrium, the fourth-order Binder cumulant of the order parameter
(i.e. the Polyakov Loop $W$)
\begin{equation}
B_{4}=\frac{\left\langle W\text{\textsuperscript{4}}\right\rangle }{\left\langle W^{\text{2}}\right\rangle ^{2}}-3.\label{Binder}
\end{equation}
Then, the shifted critical coupling $\tilde{\beta}_{c,q}$ can be
found at the single crossing (fixed-) point among multiple curves
computed for different lattice sides. Interestingly, the value of
$B_{4}$ at the critical point is unique for each universality class;
so the $YM_{2}^{3d}$ theory is predicted \cite{Svet&Yaffe} to have
$B_{4}\cong1.832$ as the $2d$ Ising model \cite{BinderCumulant_Ising2d}.
In fact, our results --- see Figure (4) --- shows that the shifted
critical coupling for $q=1.10$ is given by $\tilde{\beta}_{c,q=1.1}\approx4.581(2)$
where $B_{4}\left(\tilde{\beta}\right)\cong1.825(5),$ thus it also
agrees with predictions from universality. 

Thence, considering that $T^{-1}=a\cdot L_{t}$ and the lattice spacing
$a$ is given at leading order by $a\sim\nicefrac{1}{\beta\cdot\sqrt{\sigma}}$
\cite{Teper_3d_SUN_theories}, we may conclude that the deconfinement
critical temperature is really shifted upwards up to 30\% by nonextensive
effects when $q=1.10,$ as it was previously hypothesized. 

In this same vein, we see from compiled data in Table (1) that critical
exponents $z=2.139(9)$ and $\beta/\nu=0.124(4)$ of the $YM_{2}^{3d}$
gauge theory simulated at $q=1.10$ --- with the corrected critical
coupling $\tilde{\beta}_{c,q=1.1}\approx4.5805(8)$ --- are compatible
with computations in the BG ensemble (i.e., using $q=1.0$ and $\beta_{c}=3.4505$),
to know $z=2.008(8)$ and $\beta/\nu=0.127(5).$ Besides that, the
results nicely agree with values from literature for the critical
$2d$ Ising model in the BG ensemble \cite{Okano-Otobe-Zheng}, where
$z=2.155(3)$ and $\beta/\nu=0.125.$ These are nontrivial evidences
that the universality hypothesis among such systems \cite{Svet&Yaffe}
holds even when they are studied in different (but equivalent, see
\cite{Morishita-Johal_EGE-MUCA-Tsallis}) ensembles.

\section{Concluding remarks}

We have designed a generalized hybrid heat-bath algorithm (q-MHB)
to perform ab initio simulations of $SU\left(2\right)$ lattice gauge
fields on the nonextensive ensemble of Tsallis. The algorithm emerges
as an exact solution for a generalized detailed balance equation already
proposed in \cite{Drugo_qMetropolis_Ising}. Through group embedding
this scheme can be adapted to any gauge group $SU\left(N\right).$
Then, to verify the numerical performance of the algorithm, as a function
of $q,$ we checked the scaling $\tau_{int}=a\cdot L_{side}^{z}$
of the integrated correlation time of an extended critical plaquette.
We have observed that employing the generalized ensemble of Tsallis
with $q>1$ in association to overrelaxation allowed for improvements
on simulation performance of up to 9\%.

As discussed by Morishita \cite{Morishita-Johal_EGE-MUCA-Tsallis}
the Tsallis parameter $q$ may be physically interpretaded as the
strenght of an effective thermal coupling to a finite heat-bath. More
explicitly, by considering the heat capacity of that bath to be $C_{v}^{HB}$
and $k$ as a constant with proper dimension, one has $q=1-k/C_{v}^{HB}.$
Then, the canonical Boltzmann-Gibbs ensemble is recovered when $C_{v}^{HB}\rightarrow\infty,$
which implies the $q\rightarrow1^{-}$ limit. On the other hand, in
the oposite regime, the microcanonical ensemble of Boltzmann emerges
when $C_{v}^{HB}\rightarrow0^{+},$ i.e. when $q<0.$ Both such limits
obey $C_{v}^{HB}\geq0$ and so are said to be \textit{weakly coupled}.

The remaining mathematical possibility is to chose $C_{v}^{HB}<0$
to produce $q>1.$ This elusive regime is known as \textit{strongly
coupled}, in the sense that its thermal fluctuations are stronger
than in the canonical/microcanonical limits. From a purely computational
viewpoint it has been proved \cite{Morishita-Johal_EGE-MUCA-Tsallis}
that simulations with $q>1$ are equivalent to ones in the multicanonical
ensemble (MUCA) of Berg \cite{MUCA_Berg}. Thus, while the strongly
coupled nonextensive approach is the most efficient one in reducing
tunneling-times around phase transitions \cite{Morishita-Johal_EGE-MUCA-Tsallis,DHE_Gross},
as also corroborated by our performance analysis, the physical interpretation
of negative heat capacities of reservoirs is still debated. 

Furthermore, we have employed our generalized heat-bath algorithm
to study the short-time (relaxation) dynamics of the $SU\left(2\right)$
gauge theory in the Tsallis ensemble \cite{DTC-Frigori}. To do so,
a serie of (initially orderly) gauge configurations was prepared and
then evolved during some hundred Monte Carlo steps. During such a
temporal evolution a set of observables Eq.(\ref{M_Poly}), Eq.(\ref{U_t})
and Eq.(\ref{A_Poly}) was measured. After that, power-law scaling
relations Eq.(\ref{M_t_scaling}), Eq.(\ref{U_T_scaling}) and Eq.(\ref{A_Poly_scaling})
were carefully adjusted to data to obtain the (static and dynamic)
critical exponents collected on Table (1). By considering only the
regions with best fit-qualities $\left(\chi^{2}/dof\simeq1\right)$
we have verified that long-standing universality arguments of \cite{Svet&Yaffe}
hold for the $2d$ Ising model, in BG ensemble, and the critical $SU(2)$
gauge theory at Tsallis ensemble. Notwithstanding, to ensure such
a perfect matching, the nonextensively induced shift on the lattice
coupling $\left(\beta_{c}\rightarrow\tilde{\beta}_{c,q\neq1}\right)$
had to be precisely calculated. 

To determinate that shift of the critical gauge coupling as a function
of $q$ we have proposed a new approach based on finite-size scaling.
Here a nonequilibrium scaling relation Eq.(\ref{A_Poly_scaling})
was fitted to data while varying $\beta$ to locate the new critical
region in the Tsallis ensemble (i.e., whenever $q\neq1$). The method
so introduced was successfully compared with a traditional one, the
fourth-order Binder cumulant. In addition, both approaches agree that
deconfinement temperature is increased by about 30\% when a phenomenologically
favoured value $q=1.10$ was employed \cite{q-scalling,PHENIX}. 

Once nonextensive simulation setups analogous to ones here presented
are applicable to lattice QCD, one would expect to be able to better
describe early nonequilibrium stages of hadronic collisions from first
principles. For instance, it would be interesting to cross-check how
universal nonextensive effects relate the $3d$ $O(4)$ model ---
accessible through algorithms on section IV, and \cite{MHB} ---
and QCD. Finally, further pieces of encouragement in this direction
is that the Tsallis framework is well suited to describe systems showing
power-law relaxation in time and energy, as well as those relaxing
by nonergodic occupation of phase space due to unusual underlying
microscopic dynamics. All these peculiar features are typically found
during transient times of hadronic collisions \cite{PHENIX_AuAu,PHENIX,STAR,TBW,Why_Tsallis_in_hQCD}.

\section*{Acknowledgements}

The author thanks A. Mihara and R. da Silva for useful discussions
and UTFPR by finantial support. Numerical simulations were performed
at SGI-Altix at CENAPAD/Unicamp under project 501.

\end{document}